\documentclass[acmsmall, screen]{acmart}
\usepackage{url}
\usepackage{tcolorbox}
\usepackage{listings}
\usepackage{xcolor}
\usepackage[export]{adjustbox}
\tcbuselibrary{listings} 
\usepackage{booktabs}
\usepackage{graphicx}
\usepackage{amsmath}
\usepackage{flushend}
\usepackage{balance}
\usepackage{fancyhdr}
\usepackage{multirow}
\usepackage{hyperref}
\hypersetup{
    colorlinks=true,
    linkcolor=blue,
    filecolor=magenta,      
    urlcolor=cyan,
    citecolor=blue,
    pdfpagemode=FullScreen,
    }
\usepackage{listings}

\usepackage{xcolor}

\definecolor{codegreen}{rgb}{0,0.6,0}
\definecolor{codegray}{rgb}{0.5,0.5,0.5}
\definecolor{codepurple}{rgb}{0.58,0,0.82}
\definecolor{backcolour}{rgb}{0.95,0.95,0.92}

\lstdefinestyle{mystyle}{
  label=code:sample,
  backgroundcolor=\color{backcolour}, commentstyle=\color{codegreen},
  keywordstyle=\color{magenta},
  numberstyle=\tiny\color{codegray},
  stringstyle=\color{codepurple},
  basicstyle=\ttfamily\footnotesize,
  breakatwhitespace=false,         
  breaklines=true,                 
  captionpos=t,                    
  keepspaces=true,                 
  numbers=left,                    
  numbersep=2pt,                  
  showspaces=false,                
  showstringspaces=false,
  showtabs=false,                  
  tabsize=1
}

\usepackage[font=small,labelfont=bf]{caption}

\AtBeginDocument{%
  \providecommand\BibTeX{{%
    \normalfont B\kern-0.5em{\scshape i\kern-0.25em b}\kern-0.8em\TeX}}}

\newcommand{\ea}{\textit{et al.}}

\newcommand{\find}[1]{
\begin{tcolorbox}[leftrule=1mm,toprule=0mm,bottomrule=0mm,left=1pt,right=2pt,top=2pt,bottom=2pt]
\em #1
\end{tcolorbox}
}

\newcommand{\smallsection}[1]{\noindent {\bf \underline{#1}}.\hspace{1mm}}

\begin{document}
\title{Refining ChatGPT-Generated Code: Characterizing and Mitigating Code Quality Issues}

\thispagestyle{fancy}

\author{Yue Liu}
\email{yue.liu1@monash.edu}
\affiliation{%
  \institution{Monash University}
  \country{Australia}
}
\author{Thanh Le-Cong}
\email{congthanh.le@student.unimelb.edu.au}
\affiliation{%
  \institution{The University of Melbourne}
  \country{Australia}
}

\author{Ratnadira Widyasari}
\email{ratnadiraw.2020@phdcs.smu.edu.sg}
\affiliation{%
  \institution{Singapore Management University}
  \country{Singapore}
}

\author{Chakkrit Tantithamthavorn}
\email{chakkrit@monash.edu}
\affiliation{%
  \institution{Monash University}
  \country{Australia}
}

\author{Li Li}
\email{lilicoding@ieee.org}
\affiliation{%
  \institution{Beihang University}
  \country{China}
}

\author{Xuan-Bach D. Le}
\email{bach.le@unimelb.edu.au}
\affiliation{%
  \institution{The University of Melbourne}
  \country{Australia}
}

\author{David Lo}
\email{davidlo@smu.edu.sg}
\affiliation{%
  \institution{Singapore Management University}
  \country{Singapore}
}

\begin{abstract}
Since its introduction in November 2022, ChatGPT has rapidly gained popularity due to its remarkable ability in language understanding and human-like responses. ChatGPT, based on GPT-3.5 architecture, has shown great promise for revolutionizing various research fields, including code generation. 
However, the reliability and quality of code generated by ChatGPT remain unexplored, raising concerns about potential risks associated with the widespread use of ChatGPT-driven code generation.

In this paper, we systematically study the quality of 4,066 ChatGPT-generated code implemented in two popular programming languages, i.e., Java and Python, for 2,033 programming tasks. 
The goal of this work is three folds. First, we analyze the correctness of ChatGPT on code generation tasks and uncover the factors that influence its effectiveness, including task difficulty, programming language, time that tasks are introduced, and program size. 
Second, we identify and characterize potential issues with the quality of ChatGPT-generated code. 
Last, we provide insights into how these issues can be mitigated.
Experiments highlight that out of 4,066 programs generated by ChatGPT, 2,756 programs are deemed correct, 1,082 programs provide wrong outputs, and 177 programs contain compilation or runtime errors. Additionally, we further analyze other characteristics of the generated code through static analysis tools, such as code style and maintainability, and find that 1,930 ChatGPT-generated code snippets suffer from maintainability issues. Subsequently, we investigate ChatGPT's self-repairing ability and its interaction with static analysis tools to fix the errors uncovered in the previous step. Experiments suggest that ChatGPT can partially address these challenges, improving code quality by more than 20\%, but there are still limitations and opportunities for improvement.
Overall, our study provides valuable insights into the current limitations of ChatGPT and offers a roadmap for future research and development efforts to enhance the code generation capabilities of AI models like ChatGPT.
\end{abstract}



\maketitle

\section{Introduction}

Since launching in November 2022, ChatGPT~\cite{chatgpt}, an AI-powered chatbot developed by OpenAI, has rapidly gained popularity. Within just two months, ChatGPT had reached 100 million unique users, surpassing even the fastest-growing social network, TikTok, in user acquisition~\cite{chatgptrising}. Due to its remarkable ability in language understanding and human-like answering, ChatGPT has shown great promise in revolutionizing various research fields, including code generation due to it being trained on extensive repositories of source code~\cite{chatgpt}. Interestingly, users without any coding experience can use the model to generate code snippets from natural language requirements. Although ChatGPT's ability on code generation tasks has been \emph{informally} receiving positive feedback from the community, there exists no study that \emph{formally} investigates the reliability and quality of code generated by ChatGPT.

Despite the great promise of ChatGPT in code generation, formally and thoroughly studying the reliability and quality of code generated by ChatGPT is becoming increasingly critical. This is due to ChatGPT now being used not only by professional developers but also by novice programmers and individuals with no coding experience. 
Code quality issues in ChatGPT-generated code, if not properly identified and addressed, may unduly affect code comprehension, introduce bugs, or create security vulnerabilities in users' projects~\cite{she2023pitfalls}. 
Consequently, the widespread adoption of ChatGPT for code generation could potentially lead to a decline in the overall quality of software systems. 
Therefore, it is crucial to examine and address the common code quality issues that may arise from using ChatGPT-generated code.

In this paper, motivated by the above challenges, we are the first to formally study the reliability and quality of ChatGPT-generated code. Our objectives are (1) to analyze the correctness of ChatGPT-generated code, (2) to identify and characterize code quality issues that may arise, and (3) to examine different prompts that leverage feedback from static analysis tools and runtime errors to guide ChatGPT in mitigating code quality issues. Through experiments addressing the following three research questions, our work provides valuable insights that help increase awareness within the community regarding code quality issues in ChatGPT-driven code generation. 
\begin{itemize}
    \item \textbf{RQ1:} \textit{\textbf{(Performance)} How effective is ChatGPT on code generation for programming tasks}?
    \item \textbf{RQ2:} \textit{\textbf{(Bugs and Issues)} What are the common issues in ChatGPT-generated code?}
    \item \textbf{RQ3:} \textit{\textbf{(Repair with Prompting)} Can ChatGPT fix the code quality issues with prompting? }
\end{itemize}

To answer these questions, we first construct a benchmark dataset containing a total of 2,033 programming tasks from LeetCode, with 501 classified as easy, 1,064 as medium, and 468 as hard. 
We then evaluate the ChatGPT-generated code for these programming tasks against LeetCode's test suite to evaluate ChatGPT's performance on code generation. Next, we employ static analysis tools including Pylint~\cite{thenault2001pylint}, Flake8~\cite{cordasco2010flake8}, PMD~\cite{copeland2005pmd}, and CheckStyle~\cite{burn2003checkstyle} to examine ChatGPT-generated code. Based on feedback from static analysis tools and runtime errors, we conduct an open card sort discussion~\cite{spencer2009card} to characterize common code quality issues including compilation and runtime errors, wrong outputs, code style and maintainability, and performance and efficiency. Finally, we attempt to mitigate the identified code quality issues by using several fixing-prompts, i.e., prompts that request ChatGPT to fix issues. To do so, we experiment with fixing-prompts with and without feedback from static analysis tools and runtime errors.

Our experimental results lead to the following findings:
(1). On various code generation tasks on Python and Java, 66\% and 69\% of Python and Java programs generated by ChatGPT are functionally-correct programs, i.e., programs that pass all given test cases. We observed that the performance is attributed to various factors such as task difficulty, the time when tasks are introduced, and program size. Especially, ChatGPT's performance drops up to five times on new programming tasks introduced after January 2022, highlighting the model's limitations in adapting to new programming tasks. 
(2). We also identified that the generated code commonly suffers from different code quality issues, such as compilation and runtime errors, wrong outputs, code style and maintainability issues.
For instance, among ChatGPT-generated code that passed the test cases, 53\% of the Java code and 37\% of the Python code exhibited code style and maintainability issues. 
This highlights the importance of addressing such problems to ensure the long-term success of AI-driven code generation. In other words, developers and users still need to take appropriate measures to improve the overall quality of the ChatGPT-generated code.
(3). Our study on ChatGPT's self-repairing capabilities revealed that ChatGPT can partially fix code quality issues in the generated code with feedback from static analysis tools and runtime errors. Moreover, the effectiveness of ChatGPT in addressing code quality issues varies depending on the feedback information, programming languages, and code quality issues.

To summarize, our paper makes the following contributions:

\begin{itemize}
    \item Conduct a comprehensive study to evaluate the reliability and quality of ChatGPT-generated code;
    \item Identify and characterize common code quality issues in ChatGPT-generated code;
    \item Introduce a new time-sensitive dataset comprising 2033 programming tasks and 4066 ChatGPT-generated code snippets implemented in two popular programming languages: Java and Python, with 2,553 code with quality issues;
    \item Conduct an exploration study on ChatGPT's self-repairing capability for code quality issues.
\end{itemize}

To support the open science initiative, we publish the studies dataset and a replication package, which is publicly available at \url{ https://github.com/yueyueL/ChatGPT-CodeGenAnalysis}

\section{Background}
\label{sec:background}

\subsection{Large Language Model}
Large Language Models (LLMs) have achieved impressive performance on a wide range of natural language processing (NLP) tasks, including machine translation~\cite{radford2019language, chowdhery2022palm, hendy2023good}, summarization~\cite{radford2019language, feng2020codebert, goyal2022news}, sentiment analysis~\cite{xu-etal-2020-dombert, zhang2020sentiment}, and question-answering~\cite{radford2019language, openai2023gpt4}. 
These models, typically based on deep learning architectures such as transformers, are trained on massive amounts of text data, allowing them to learn complex language patterns and structures. By capturing both the syntax and semantics of human language, LLMs have been successful in generating coherent and contextually relevant text.

One prominent example of an LLM is ChatGPT, developed by OpenAI and based on the GPT-3 architecture. ChatGPT demonstrates an unprecedented ability to understand and generate human-like text, making it well-suited for a variety of applications, including code generation. By training ChatGPT on extensive source code repositories, the model has become capable of generating code snippets and solving programming problems with remarkable accuracy~\cite{hou2023large}.

\subsection{Motivation}

While LLMs have shown great promise in code generation, the reliability of the generated code is questionable. The problem has become more critical with the emergence of ChatGPT, as LLMs-driven code generation is now being used not only by experienced developers but also by novice programmers or even individuals with no coding experience, who may be unaware of the code quality issues. 


\begin{figure}
    \centering
    \includegraphics[width=\columnwidth]{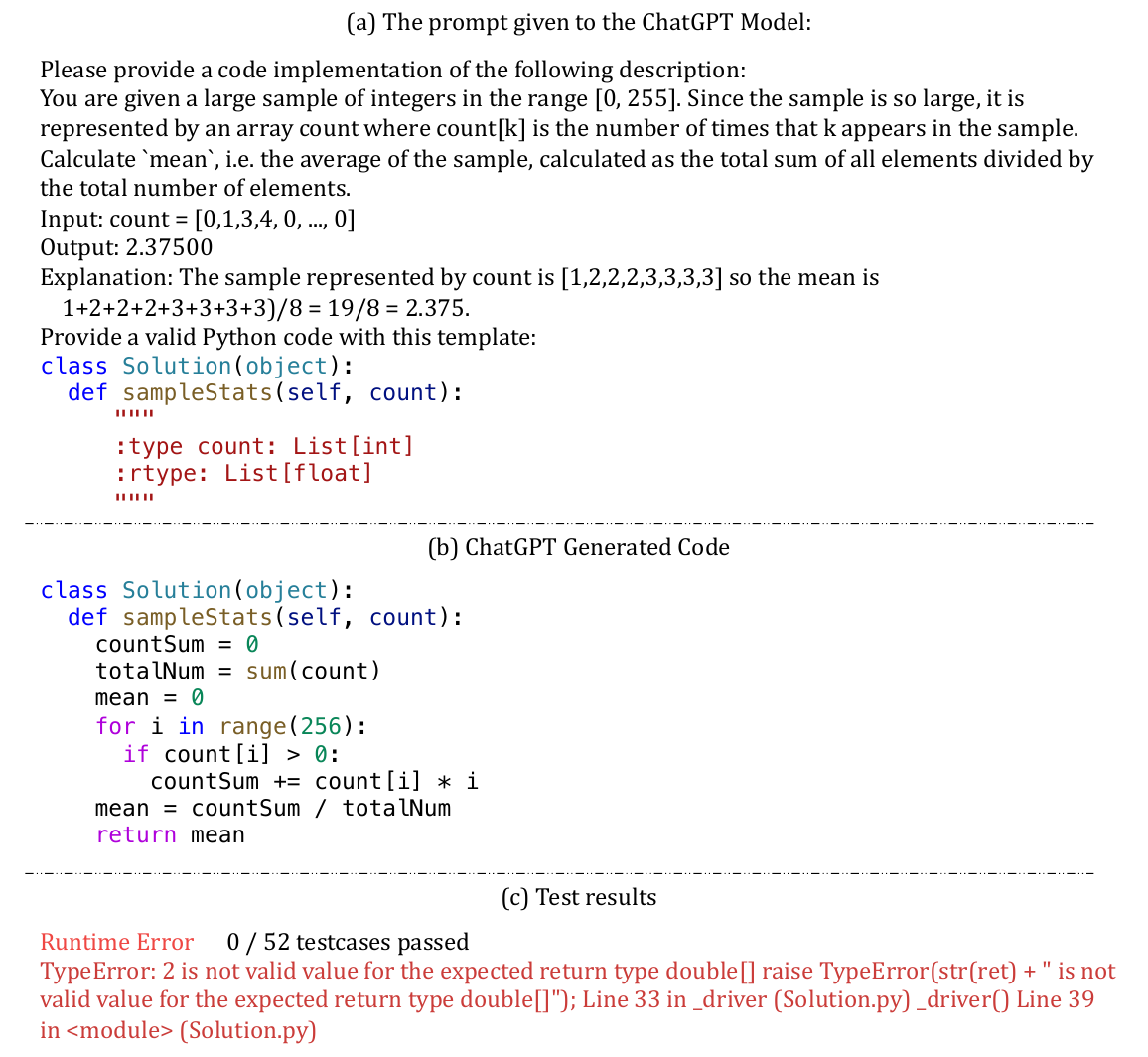}
    \caption{Example of an buggy code generated by ChatGPT for solving the LeetCode Problem 1093 - 'Statistics from a Large Sample'}
    \label{fig:bug_example}
\end{figure}

Figure~\ref{fig:bug_example} describes a motivating example for our study.
Figure~\ref{fig:bug_example}(a) presents the prompts to ChatGPT, which combine the task description, constraints, and predefined code templates.
The programming task is called “Statistics from a Large Sample” \cite{leetcode1093}.
The problem requires ChatGPT to generate a code that calculates the mean of a large sample of integers, represented by a \texttt{count} array where \texttt{count[k]} represents the frequency of integer k in the sample.
Figure~\ref{fig:bug_example}(b) presents buggy code generated by ChatGPT to solve this problem. While looking straightforward and correct, the ChatGPT-generated code produces the incorrect output from the example test, as shown in Figure~\ref{fig:bug_example}(c).
The expected output from the test is 2.375, while the result from ChatGPT-generated code is 2. 
The root cause is that \texttt{mean} is calculated using integer division (rounding down to an integer) since both \texttt{countSum} and \texttt{totalNum} are integers.
Though the error is quite simple, it can be difficult for developers or programmers who are not familiar with Python programming languages to detect. It can also lead to more complex errors in other functions that call to this function without the awareness of the error. 

\begin{lstlisting}[language=Python, caption=A code smell generated by ChatGPT for solving the LeetCode Problem 1838 - 'Frequency of the Most Frequent Element', label=fig:smell_example]
def getMinDistance(self, nums: List[int], target: int, start: int) -> int:
    min_diff = float('inf')
    min_index = -1
    for i in range(len(nums)):
        if nums[i] == target:
            diff = abs(i - start)
            if diff < min_diff:
                min_diff = diff
                min_index = i
    return min_diff
\end{lstlisting}
Additionally, we also observed that the quality of the ChatGPT-generated code may still be poor even if it is functionally correct. Code~\ref{fig:smell_example} illustrates an example of poor-quality code generated by ChatGPT. This is a simplified version of code generated by ChatGPT for LeetCode Problem 1838, `Frequency of the Most Frequent Element'. The \texttt{min\_index} variable is declared on line 3 and assigned values on line 9, but it is never used elsewhere in the code. This is a minor code smell, but it is worth noting that this issue occurs in a simple 10-line code for a common problem. Let's imagine complex tasks and code, could we ensure that ChatGPT-generated code does not contain smells, bugs, or even vulnerabilities? This realization motivated us to conduct a comprehensive study on the quality issues present in ChatGPT-generated code. Our study aims to not only enhance our understanding of these issues but also to provide suggestions for mitigating them.

\begin{figure}
    \centering
    \includegraphics[width=\columnwidth]{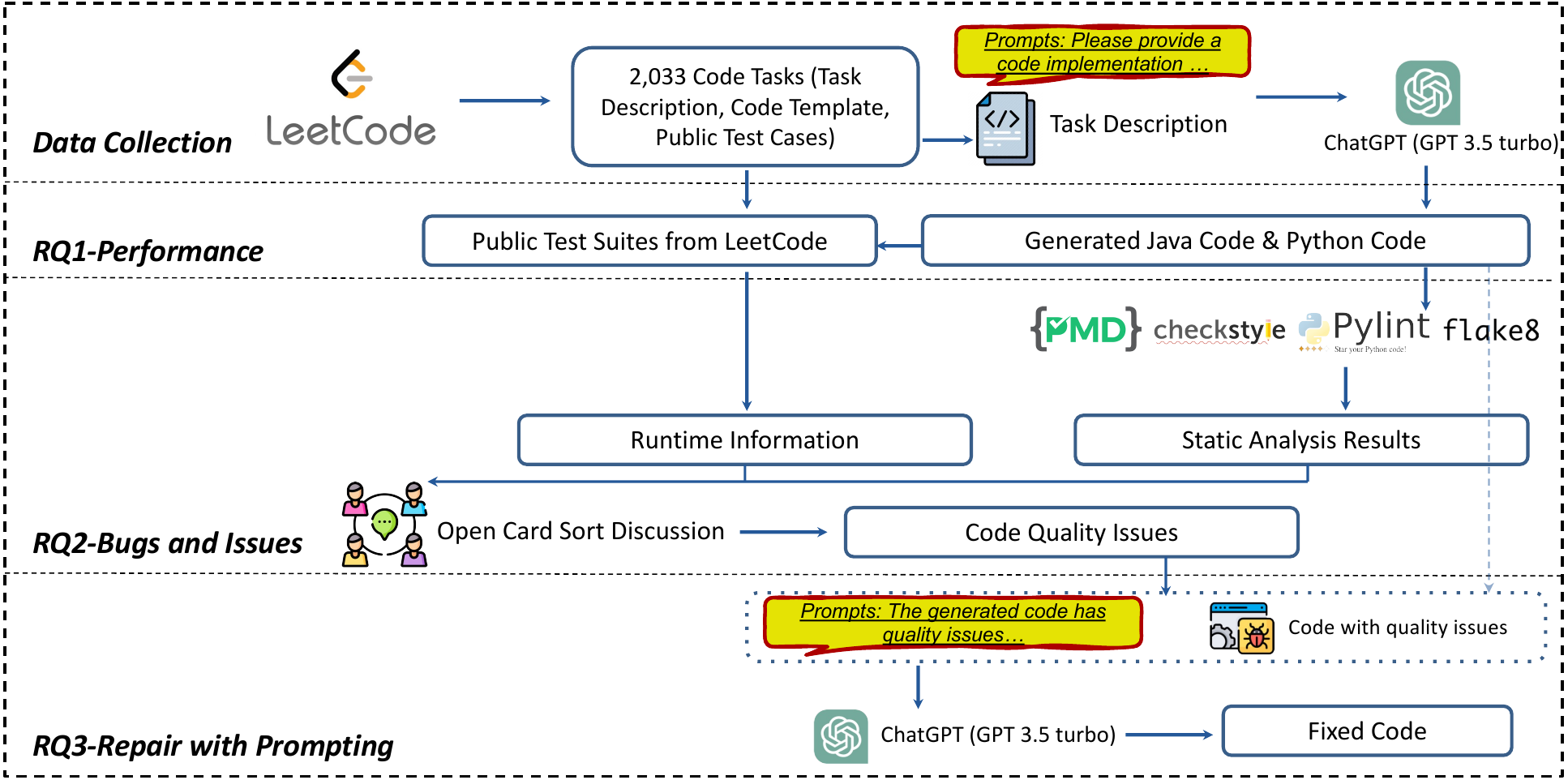}
    \caption{Overview of our workflow}
    \label{fig:overview}
\end{figure}

\section{Study Setup}
\label{sec:approach}
In this section, we present the comprehensive setup of our empirical study.
We describe the research questions, illustrate the workflow of our study design, and provide an in-depth description of the benchmark dataset construction and analysis. Furthermore, we detail the characteristics of the ChatGPT model employed in this study.

\subsection{Research Design}
In this empirical study, we aim to answer the following research questions.

\vspace{1mm}

\noindent\textbf{RQ1.} \textit{How effective is ChatGPT on code generation for programming tasks?} Despite informally receiving positive feedback from the community, there is a lack of comprehensive study on the performance of ChatGPT in code generation. This research question aims to measure how well ChatGPT could generate code for programming tasks and to analyze the factors that impact its performance, including task complexity, difficulty, and time that tasks are introduced.

\noindent\textbf{RQ2.} \textit{What are the common issues in ChatGPT-generated code?}  This research question aims to analyze issues in ChatGPT-generated code using popular static analysis tools and categorize them into common categories.

\noindent\textbf{RQ3.} \textit{Can ChatGPT fix the code quality issues with prompting?} Conversational AI models such as ChatGPT allow users to provide feedback to allow ChatGPT to revise its output. This research question aims to investigate whether ChatGPT can correct coding issues based on runtime errors, feedback from the compiler, and static analysis tools.

\vspace{1mm}

Figure~\ref{fig:overview} presents the comprehensive workflow of our study, outlining the steps taken to answer the above research questions. 
Our approach starts with a data collection stage, where we collect 2,033 programming tasks from LeetCode.
These tasks, including task descriptions, code templates, and public test cases, serve as the foundation for our research.
Subsequently, ChatGPT is prompted to generate code solutions in Java and Python for these tasks. 
The generated code is then evaluated for performance based on task-specific test cases to address RQ1. 
This evaluation allows us to assess the effectiveness of ChatGPT in code generation, considering various dimensions such as task complexity and programming language types.
For all the generated code, we also employ automated static analysis tools including PMD~\cite{copeland2005pmd} and Checkstyle~\cite{burn2003checkstyle} for Java, and Pylint~\cite{thenault2001pylint} and Flake8~\cite{cordasco2010flake8} for Python. 
These tools enable us to identify and categorize code quality issues systematically. 
Combining the static analysis results with runtime information provided by compilers, we engage in a discussion using open card sorting. 
Through classifying identified bugs and issues, this systematic approach provides comprehensive answers to RQ2.
The final stage involves the repair of code quality issues (RQ3), where ChatGPT, upon receiving targeted prompts, attempts to repair the faults. These prompts are based on feedback from both static analysis tools and runtime error messages. This stage is important in determining ChatGPT's ability to self-repair and improve the code based on conversational AI feedback mechanisms. It provides insights into the practical application of ChatGPT in real-world coding scenarios, where iterative feedback and correction play a significant role.

\subsection{Constructing Benchmark Dataset}
Existing benchmarks for evaluating AI-based code generation are often limited and outdated.
Specifically, popular benchmarks like HumanEval~\cite{chen2021evaluating} encompassing 164 Python tasks, and MBPP~\cite{austin2021program} containing 974 programming tasks, have been widely used by prior research~\cite{chen2021evaluating, chen2022codet, liu2023your, cassano2023multipl}.
However, they were released prior to 2021 and lack detailed temporal metadata for the tasks.
Therefore, such small and outdated datasets are not ideal for evaluating modern generative models like ChatGPT, since they lack diversity and may have been used in the training data of modern AI models, thus providing unrealistic performance evaluation for these models.
To address this issue, Fan~\ea~\cite{fan2022automated} introduce a new dataset, LMDefects, that contains 113 Java programming tasks released after Jun 2021. The dataset was collected from LeetCode, a well-known online platform that offers a variety of coding challenges to help programmers enhance their abilities and prepare for technical interviews. 
The dataset, however, is still relatively small and focused solely on Java programming tasks.

\begin{figure}[t]
  \centering
  \begin{minipage}{0.5\linewidth}
    \centering
    \includegraphics[height=3.6cm, keepaspectratio]{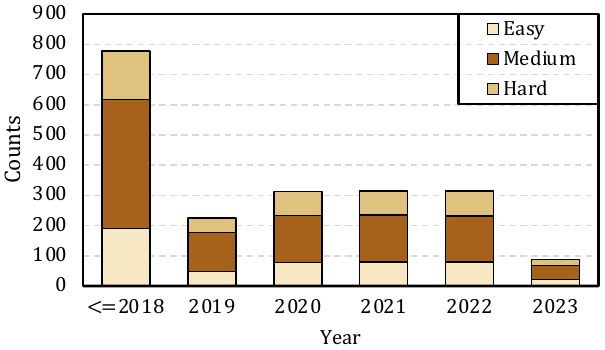} 
    \caption{Task distribution across time}
    \label{fig:task_dis_time}
  \end{minipage}%
  \hfill
  \begin{minipage}{0.5\linewidth} 
    \centering
    \includegraphics[height=3.6cm, keepaspectratio]{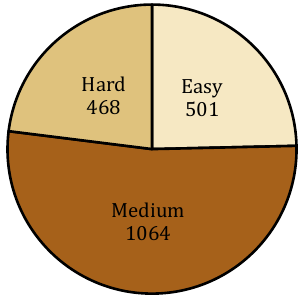}
    \caption{Task distribution across difficulty }
    \label{fig:task_accross_diff}
  \end{minipage}
\end{figure}

In this study, we extend LMDefects by collecting all accessible programming tasks and the relevant official public test suites in LeetCode, and investigate ChatGPT's ability in generating code in both Java and Python.
At the time of data collection (March 2023), there were 2,617 task problems available on LeetCode. 
These problems cover various topics, including data structures, algorithms, databases, and concurrency.
For our dataset, we focused on the problems that were designed specifically for Java and Python, as these two languages are widely used and have a large community of developers.
Additionally, in order to provide a fair and accessible dataset, we filtered out the premium tasks that require a subscription to access.
After this filtering process, we successfully collected 2,033 programming tasks from LeetCode.
For each task listed on LeetCode, we collected a comprehensive set of data including the task description, example test cases, constraints, and predefined code templates for both Python and Java.
Figure~\ref{fig:task_dis_time} and Figure~\ref{fig:task_accross_diff} present the distribution of tasks across time and difficulty levels classified by LeetCode.
As shown in Figure~\ref{fig:task_dis_time}, while most tasks are from before 2021, there are still more than 400 test cases for evaluating ChatGPT's code generation capabilities. 
This temporal diversity is important for a fair evaluation of the model's performance over different periods.
Figure~\ref{fig:task_accross_diff} shows that out of the 2,033 tasks in our dataset, we found that 501 were classified as easy, 1,064 as medium, and 468 as hard.

\subsection{The ChatGPT Model}
ChatGPT is a large language model that can provide detailed responses given natural language instructions. We use the model that has been fine-tuned from the GPT3.5 models~\cite{gpt35} using Reinforcement Learning from Human Feedback~\cite{stiennon2020learning}. In our study, we used the ChatGPT-March-23 version~\cite{chatgptrelease}, which was trained on data up to 2021. To instruct ChatGPT, we followed the zero-shot prompt setting which does not include examples of code generation in the prompt as illustrated in Code~\ref{fig:prompt}. To mitigate the randomness of ChatGPT, we make ChatGPT deterministic by setting the temperature to 0 and running the model once for each task, using the first generated output for evaluation.

\begin{lstlisting}[ caption=Prompt Template, label=fig:prompt]
Please provide a code implementation of the following description:
<description of a programming task>
Provide a valid <programming language> code with this template:
<solution template provided containing the input and output specifications>
\end{lstlisting}

\begin{table}[t]
  \centering
  
  \caption{Zero-shot pass rate accuracy on LeetCode}
    \begin{tabular}{lcccc}
    \toprule
    \textbf{pass@1} & \textbf{Easy} & \textbf{Medium} & \textbf{Hard} & \textbf{Overall} \\
    \midrule
    \textbf{Python} & 0.890 & 0.674 & 0.400 & 0.664 \\
    \textbf{Java} & 0.860 & 0.710 & 0.468 & 0.691 \\
    \midrule
    \textbf{P-value} & 0.346 & 0.654 & 0.535 & 0.471 \\
    \textbf{Effect Size} & 0.015 & -0.012 & -0.005 & 0.002 \\
    \bottomrule
    \end{tabular}%
  \label{tab:rq1acc_leetcode}%
\end{table}%

\section{RQ1: Performance}
\label{sec:RQ1}

\smallsection{Experimental Design}
In this section, we present the results for RQ1, which investigates the effectiveness of ChatGPT in code generation. 
To mitigate the randomness of ChatGPT, we make ChatGPT deterministic by setting the temperature to 0 and running the model once for each task, using the first generated output for evaluation.
ChatGPT's performance is measured with zero-shot pass-rate (\textit{pass@1}), which measures whether the model produces a correct solution (i.e., passes all the test cases) on the first attempt.
For example, if ChatGPT generates code snippets for 10 tasks and 7 of them pass the test cases in the first attempt, the pass@1 accuracy would be 0.70.
We also conducted the Mann-Whitney U rank test~\cite{mann1947test} to measure the statistical significance of the performance differences by ChatGPT across factors. 
The Mann-Whitney U rank test is a non-parametric statistical test used to compare two independent samples to determine whether there is a significant difference between the two distributions, while the Cliff's Delta~\cite{macbeth2011cliff} effect size measures the magnitude of the difference between the samples.

\smallsection{Result}
Table~\ref{tab:rq1acc_leetcode} presents the pass rate of ChatGPT for LeetCode tasks with different difficulties in both Python and Java. 
It can be seen that ChatGPT performs better on easy tasks than on medium and hard tasks. 
For Python, the model achieves a pass@1 accuracy of 0.890 for easy tasks,  indicating that ChatGPT can handle 89\% of easy tasks in one attempt. 
However, the performance drops to 0.674 for medium tasks and further decreases to 0.400 for hard tasks.  
Similarly, for Java, the model attains a pass@1 accuracy of 0.860 for easy tasks, 0.710 for medium tasks, and 0.468 for hard tasks. 
These findings suggest that the difficulty level of tasks has a significant impact on the performance of ChatGPT in code generation.
Table~\ref{tab:rq1acc_leetcode} also shows the results from the Mann-Whitney U test on performance differences between Python and Java.
Although ChatGPT performs slightly better in Java for medium ($\uparrow 5.3\%$) and hard tasks ($\uparrow 17\%$), their difference in performance is not significant with a p-value of at least 0.53 and an effect size value less than 0.02~\cite{macbeth2011cliff}.



\begin{table}[t]
  \centering
  \caption{Effect Sizes and P-values for Pass vs Fail Comparisons in Python and Java}
  \scalebox{0.9}{
    \begin{tabular}{cccr}
    \toprule
    \textbf{Comparison (@pass \textit{vs} @fail)} & \textbf{Language} & \textbf{P-value} & \textbf{Effect Size (Cliff's Delta)} \\
    \midrule
    \multirow{2}{*}{Time period} & Python & $<$0.001 & 0.511 \\
          & Java  & $<$0.001 & 0.446 \\
    \multirow{2}{*}{Program length} & Python & $<$0.001 & 0.249 \\
          & Java  & $<$0.001 & 0.309 \\
    \bottomrule
    \end{tabular}%
    }
  \label{tab:rq1_comparism_pass_fail}%
\end{table}

\begin{figure}[t]
    \centering
    \includegraphics[width=0.6\textwidth]{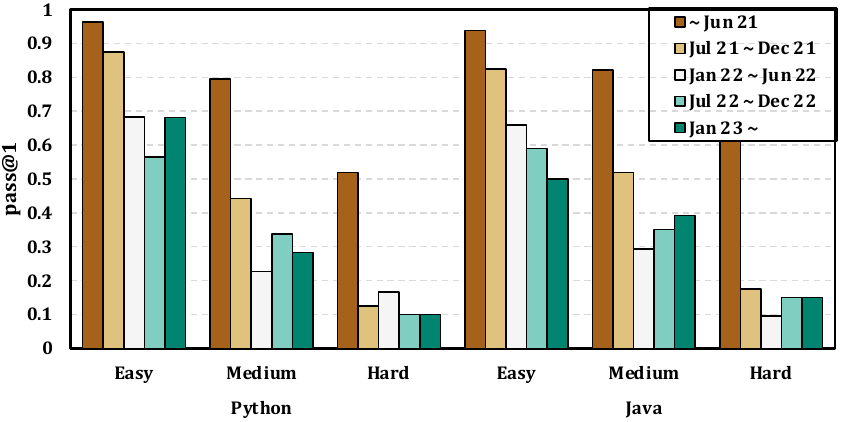}
    \caption{Pass Rate by Difficulty and Time Period}
    \label{fig:rq1time}
\end{figure}

As ChatGPT (GPT-3.5-turbo) is trained solely on data until September 2021~\cite{chatgpt}, it's also important to measure its performance changes as new challenges arise.
Figure~\ref{fig:rq1time} illustrates the pass rates of ChatGPT across different difficulty levels (easy, medium, and hard) and programming languages (Python and Java) over five distinct time periods. 
The chart shows that the performance of ChatGPT declines over time for both Python and Java.
Specifically, ChatGPT can solve more than half of the hard-level code tasks before June 2021, but its performance reduces drastically to nearly 0.1 for the subsequent time periods.
The decline in performance is not as pronounced for easy-level tasks, which indicates that ChatGPT still maintains some level of proficiency when dealing with simpler problems, even as time progresses. 
As shown in Table~\ref{tab:rq1_comparism_pass_fail}, the Mann-Whitney U test indicates that the time period when tasks are introduced has a statistically significant difference between passed code and failed code (p-value \textless~0.001) with a large Cliff's delta effect size.
However, this observation also highlights the model's limitations in adapting to the intricacies and nuances of more complex, newer programming challenges.
Moreover, the drop in performance of ChatGPT could be explained by a data leakage issue in which the LeetCode problem may be contained in ChatGPT's training data. Therefore, the performance of ChatGPT on old programming tasks which is published before December 2021 may only reflect the memorization capability~\cite{carlini2023quantifying} of ChatGPT instead of its real performance. Therefore, the results also highlight the need to evaluate the model on the newly-introduced dataset after September 2021 for fair evaluations. 


\begin{figure}[t]
    \centering
    \includegraphics[width=0.6\textwidth]{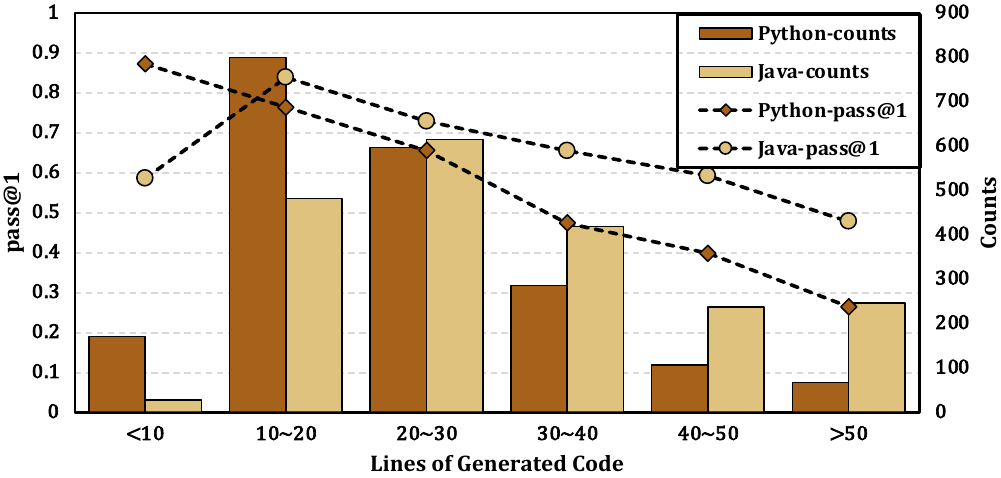}
    \caption{Pass Rate by 
    Length of Generated Program}
    \label{fig:rq1leght}
\end{figure}

In addition to difficulty levels and time periods, another factor that may impact the performance of ChatGPT is the length of the generated code. 
Figure~\ref{fig:rq1leght} presents the pass rates of ChatGPT for both Python and Java programming languages, grouped by the number of lines in the generated code. 
It is worth noting that the distribution of code lengths is not uniform, with the majority of generated code snippets falling into the 10-20 lines range for Python and the 20-30 lines range for Java. 
This discrepancy highlights the differences in verbosity and structure between the two programming languages, which might also contribute to the variations in ChatGPT's performance across different length categories.
In Figure~\ref{fig:rq1leght} there is a clear trend of decreasing the pass rate for both Python and Java, as the length of the generated code increases.
For Python, the pass@1 rate starts at 0.872 for code snippets with less than 10 lines and gradually decreases to 0.265 for code snippets with more than 50 lines. 
For Java, the pass@1 rate gradually decreases from 0.838 for code snippets with 10--20 lines to 0.478 for code snippets with more than 50 lines. 
This trend suggests that ChatGPT's ability to generate correct and bug-free code is inversely proportional to the size of the generated code. 
This could be due to the increased complexity, and the greater number of potential interactions between code elements as the code size grows, making it harder for the model to generate a correct and complete solution.
As shown in Table~\ref{tab:rq1_comparism_pass_fail}, the Mann-Whitney U test confirms the significance of the differences (p-value $<$ 0.01) with a small to medium effect size.
Overall, these findings suggest that improving the model's ability to generate longer and more complex code snippets is a valuable direction for future research and development.



In summary, our results indicate that the model's performance declines over the difficulty level and time period of code tasks.
Furthermore, the model's ability to generate correct and bug-free code is inversely proportional to the size of the generated code, suggesting that the increased complexity of longer code snippets poses a significant challenge for the model.
Based on these findings, it is recommended that future research and development efforts focus on improving the model's ability to handle more complex tasks, adapt to new programming challenges, and generate longer and more intricate code snippets. 

\find{
\textbf{Finding 1:} The performance of ChatGPT is significantly and substantially affected by task difficulty,  time that tasks are introduced, program size and programming languages.
}

\section{RQ2: Bugs and Issues}
\label{sec:RQ2}

\subsection{Static Analysis}

\smallsection{Experimental Design}
To address RQ2, our first step is to gather output from LeetCode for ChatGPT-generated code.
If the generated code does not pass the tests, we label it as ``Wrong Outputs'' to indicate failure to meet the problem requirements. 
However, passing test cases alone does not guarantee that the code is free from quality issues.  
Therefore, to further investigate the code quality and identify potential bugs, style issues, and other concerns that might impact the overall quality, we employ static analysis tools tailored for each programming language.

For Java code samples, we use PMD~\cite{copeland2005pmd} and Checkstyle~\cite{burn2003checkstyle}.
PMD is a well-known static analysis tool that inspects Java source code to identify potential problems and provides suggestions for improvements~\cite{wattanakriengkrai2020predicting}.
Checkstyle, on the other hand, statically checks Java code against a specified set of coding conventions.
Both tools evaluate the Java source code against a set of rules, reporting warnings for any violations, their priority, and the corresponding lines in the file.
Similarly, for Python code samples, we utilize Pylint~\cite{thenault2001pylint} and flake8~\cite{cordasco2010flake8}.
Pylint is a popular and comprehensive static analysis tool that enforces coding standards and detects various types of issues in Python code~\cite{siddiq2022empirical,vassallo2020configuration}. Flake8 is another widely-used tool for Python, which combines PyFlakes, pycodestyle, and McCabe to check for syntax errors, style issues, and code complexity, respectively.
These tools enable us to assess the code quality from multiple dimensions, beyond just functional correctness.

\begin{figure}[t]
    \centering
    \includegraphics[width=0.8\textwidth]{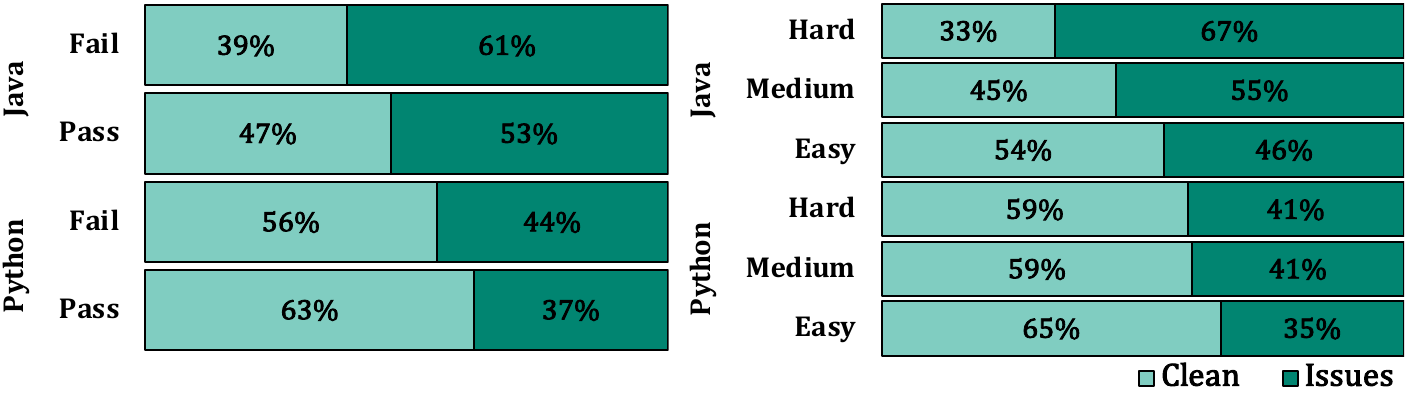}
    \caption{Code Quality Distribution by Difficulty and Language for Passed and Failed Tasks}
    \label{fig:rq2_overall_distribution}
\end{figure}

After running the static analysis tools, we gather issues found for each ChatGPT-generated program identified by the compilers and the static analysis tools.
In order to simplify the analysis and reduce the impact of false positives, we focus on more significant aspects of code quality and functionality. 
Therefore, we choose to ignore messages related to style issues such as whitespace, newline, and invalid naming conventions, which is consistent with the approach taken in prior work~\cite{siddiq2022empirical}.

\smallsection{Result}
Figure~\ref{fig:rq2_overall_distribution} presents the distribution of code quality based on the difficulty levels and programming languages for both passed and failed tasks.
The figure highlights the proportion of clean code, which refers to the code snippets without issues identified by the static analysis tools, and the code with issues.
Figure~\ref{fig:rq2_overall_distribution} shows that the proportion of clean code is generally higher for passed tasks compared to failed tasks.
For Python, 63\% of the passed tasks have clean code, while only 56\% of the failed tasks are clean. 
In the case of Java, 47\% of the passed tasks have clean code, as opposed to 39\% for failed tasks.
Additionally, it is evident that the percentage of clean code decreases as the difficulty level increases for both Python and Java. 
For example, the percentage of clean Java code decreases from 54\% for easy tasks to 45\% for medium tasks, and further drops to 33\% for hard tasks.
These findings underscore the importance of addressing code quality concerns in tandem with functional correctness to better support developers in handling complex programming tasks across different languages and domains.

\find{
\textbf{Finding 2:} Code quality issues commonly happen in both code that pass or failed test cases, highlighting the need for characterizing and addressing these concerns alongside functional correctness.
}

\subsection{Open Card Sorting Discussion}

\smallsection{Experimental Design}
To gain a deeper understanding of the common issues and patterns found in the ChatGPT-generated code, we conducted a qualitative analysis using open card sorting. 
The open card sorting process has been used in many previous studies~\cite{wan2017bug,bao2016android,lo2015practitioners} to generate categories or taxonomy from data.
In this study, we follow the card sorting process highlighted by previous studies~\cite{spencer2009card,lo2015practitioners}, which mainly consisted of two phases.
First, the preparation phase where we created cards for each programming task. This card is filled with the title of the programming task, generated code by ChatGPT, the test results, and the static analysis tools result. The second phase is the execution phase where we choose the representative random sample of 154 programming tasks for both Java and Python from a total of 2,033 programming tasks (with a 99\% confidence level and 10\% margin of error).
We analyzed and discussed each card, and iteratively sort them into groups based on their issues. During the card sorting process, we found that many of the cards could be placed into several different categories, as there can be more than one issue coming up in a given code. Once all the cards were placed into categories, we created category names based on the issue patterns that we observed.

\smallsection{Result}
After a thorough analysis and discussion of the card sorting process, the first three authors identify four different categories of issues in the ChatGPT-generated code: 

\textbf{Compilation and Runtime Errors:} 
Compilation and Runtime errors encompass issues that prevent the correct execution of the code.
These errors can arise due to various factors, such as incorrect use of a programming language's syntax, improper handling of data structures, invalid input, or exceeding the bounds of an array. 
The errors often lead to failures during compilation or runtime, and they need to be resolved before the program can function as intended.
Code~\ref{code:syntax_example} demonstrates a compilation error that occurs when ChatGPT attempts to use the \textasciicircum (bitwise XOR) operator with incompatible operand types, causing a compilation error.

\begin{lstlisting}[language=java, caption=An example of compliation error (LeetCode Problem 2564 - Java), label=code:syntax_example]
if (prefix[mid] ^ (left == 0 ? 0 : prefix[left - 1]) > queries[i][1]) {
    r = mid - 1;
} else {
    l = mid;
}
//Compiler: Solution.java:1: error: bad operand types for binary operator '^'
\end{lstlisting}

\textbf{Wrong Outputs: } Wrong Outputs represent issues in the code that cause it to produce incorrect results or fail to meet the problem requirements. 
These errors can stem from incorrect algorithms, improper handling of edge cases, or other inaccuracies in the desired logic. 
These errors can occur even when the code is syntactically correct and free from any runtime errors.
Code~\ref{code:incorrect_example} presents an example where ChatGPT provided an inaccurate solution to LeetCode Problem 746, ``Min Cost Climbing Stairs".
The issue arises due to the incorrect construction of the loop and final return statement. 
In this specific example, the input is [10, 15, 20]. 
The expected output is 15, achieved by climbing the steps with costs of 10 and 20, while skipping the step with a cost of 15. 
However, the ChatGPT-generated code produces an output of 25. 
This error occurs because the loop iterates one step more than necessary, causing the last step's cost to be included in the calculation even when it should not be.

\begin{lstlisting}[language=python, caption=A example of Wrong Outputs (LeetCode Problem 746 - Python), label=code:incorrect_example]
def minCostClimbingStairs(self, cost: List[int]) -> int:
    n = len(cost)
    dp = [0] * (n+1)
    dp[0] = cost[0]
    dp[1] = cost[1]
    for i in range(2, n+1):
        dp[i] = min(dp[i-1], dp[i-2]) + cost[i-1]
    return min(dp[n-1], dp[n])
#Compiler: AssertionError: Input: cost = [10, _15_ ,20]  Expected output: 15
\end{lstlisting}

\textbf{Code Style and Maintainability:} This category encompasses issues related to the code's formatting, organization, and adherence to established style guides.
Poorly styled or organized code is difficult to read, understand, and maintain.
Examples include inconsistent indentation, overly long lines, or lack of meaningful variable names.
Code 5 presents an example where ChatGPT provided an smelly solution to LeetCode Problem 1297, “Maximum Number of Occurrences of a Substring".
In this case, the static analysis tool PMD identified an issue on line 1: the variable `maxSize` is declared but never used.

\begin{lstlisting}[language=python, caption=A example of Code Style and Maintainability (LeetCode Problem 1297 - Python), label=code:code-smell]
def maxFreq(self, s: str, maxLetters: int, minSize: int, maxSize: int) -> int:
    count = defaultdict(int)
    res = 0

    for i in range(len(s) - minSize + 1):
        substring = s[i: i + minSize]
        if len(set(substring)) <= maxLetters:
            count[substring] += 1
            res = max(res, count[substring])

    return res
#PMD: 1297-maximum-number-of-occurrences-of-a-substring.py:1:61: W0613: Unused argument 'maxSize' (unused-argument)
\end{lstlisting}

\textbf{Performance and Efficiency:} 
Performance and efficiency issues arise when the code is not optimized for execution speed or resource usage.
These issues can lead to slow or resource-intensive programs, which can negatively impact the user experience or system stability. 
Examples include inefficient algorithms, unnecessary memory allocations, or redundant calculations. 
For example, when using ChatGPT to solve LeetCode Problem 1982, titled "Find Array Given Subset Sums", the compiler outputs a ``TIMEOUT" error due to inefficient loop control within the generated code.

\find{
\textbf{Finding 3:} Issues in ChatGPT-generated code can be categorized into four categories:  Compilation \& Runtime Errors, Wrong Outputs, Code Style \& Maintainability, Performance \& Efficiency.
}

\subsection{Quantitative Analysis}
In order to gain a comprehensive understanding of the issues present in the ChatGPT-generated code, we perform a quantitative analysis on the categorized issues identified in the open card sorting discussion. 
This analysis aims to provide insights into the frequency, distribution, and nature of the issues across different difficulty levels and programming languages.
From the card sorting results, we derive rules to classify the issues in the generated code, allowing us to identify areas where ChatGPT performs well and aspects that require improvement.
It is important to note that one generated code snippet may contain multiple issues, which can further affect the analysis.
By highlighting these issues, our analysis can guide future research and development efforts to enhance the code generation capabilities of ChatGPT and similar AI models.

\begin{table}[t]
    \centering
    \caption{Distribution of issues across difficulty levels and programming languages. P and J denotes Python and Java, respectively.}
    \label{tab:rq2_issues_distribution}
    \scalebox{0.65}{
    \begin{tabular}{lccccccccc}
    \toprule
          & \multicolumn{2}{c}{\textbf{Easy (501)}} & \multicolumn{2}{c}{\textbf{Medium (1064)}} & \multicolumn{2}{c}{\textbf{Hard (468)}} & \multirow{2}[3]{*}{\textbf{Pass (2756)}} & \multirow{2}[3]{*}{\textbf{Fail (1310)}} & \multirow{2}[3]{*}{\textbf{Sum}} \\
\cmidrule{2-7}          & \textbf{P} & \textbf{J} & \textbf{P} & \textbf{J} & \textbf{P} & \textbf{J} &       &       &  \\
    \hline
    \textbf{Compilation and Runtime Error } & 7 (1\%) & 8 (2\%) & 37 (3\%) & 32 (3\%) & 46 (10\%) & 47 (10\%) & 0 (0\%) & 177 (14\%) & 177 (4\%) \\
    \textbf{Wrong Outputs} & 47 (9\%) & 60 (12\%) & 290 (27\%) & 260 (24\%) & 229 (49\%) & 196 (42\%) & 0 (0\%) & 1082 (83\%) & 1082 (27\%) \\
    \textbf{Code Style and Maintainability} & 174 (35\%) & 230 (46\%) & 431 (41\%) & 588 (55\%) & 194 (41\%) & 313 (67\%) & 1243 (45\%) & 687 (52\%) & 1930 (47\%) \\
    \textbf{Performance and Efficiency} & 1 (0\%) & 2 (0\%) & 20 (2\%) & 16 (2\%) & 6 (1\%) & 6 (1\%) & 0 (0\%) & 51 (4\%) & 51 (1\%) \\
    \bottomrule
    \end{tabular}%
    }
\end{table}

\subsubsection{Overall Analysis}
Table~\ref{tab:rq2_issues_distribution} presents the distribution of the four issues across task difficulty levels and programming languages. 
From the table, it is evident that Compilation \& Runtime Errors and Performance \& Efficiency issues are relatively less frequent, indicating that ChatGPT is generally successful in generating syntactically correct and efficient code.
However, Wrong Output and Code Style \& Maintainability issues are more prevalent and tend to be the most common challenges faced by the generated code.
Specifically, 1,082 out of 4,066 generated code snippets (i.e., 27\%) exhibit wrong output, while 1,930 out of 4,066 (i.e., 47\%) encounter issues related to code style and maintainability.
Furthermore, as the difficulty level of the tasks increases, the prevalence of these issues also tends to rise. 
For example, 7 out of 501 generated code snippets (i.e., 1\%) for easy Python tasks exhibit compilation and runtime errors, while the number of execution errors increases to 46 out of 468 (i.e., 10\%) for hard Python tasks.
Table~\ref{tab:rq2_issues_distribution} also presents the distribution of issues across generated code that passed or failed test cases.
Out of 2,756 programs that passed all test cases, 1,243 (45\%) have issues related to code style and maintainability despite executing correctly. 
Regarding the 1,310 code that failed test cases, 14\% encounter compilation and runtime errors, 83\% exhibit wrong outputs, 4\% exhibit performance or efficiency issues, and notably, 52\% exhibited issues related to code style and maintainability on top of their functional errors.
These findings indicate that ChatGPT, while powerful, has room for improvement in automated code generation to deliver more reliable and effective AI-generated code.

\find{
\textbf{Finding 4:} Wrong Outputs and Code Style \& Maintainability issues are the most common challenges faced by the ChatGPT-generated code while Compilation \& Runtime Errors and Performance \& Efficiency issues are relatively less prevalent.
}

\begin{table*}[t]
  \centering
  \caption{Comparison of Common Compilation and Runtime Error Categories in Java and Python Programs}
    \scalebox{0.75}{
    \begin{tabular}{llrr}
    \toprule
    \textbf{Category} & \multicolumn{1}{c}{\textbf{Description}} & \multicolumn{1}{c}{\textbf{Java Count}} & \textbf{Python Count} \\
    \midrule
    \textbf{Division by Zero} & Attempt to divide by zero & 3     & 3 \\
    \textbf{Illegal Index} & Accessing an array or list with an invalid index & 45    & 25 \\
    \textbf{Concurrent Modification} & Modifying a collection during iteration & 1     & 1 \\
    \textbf{Empty Collection Access} & Accessing an element from an empty collection & 2     & 3 \\
    \textbf{Key Not Found} & Accessing a non-existent key in a dictionary or map & 1     & 13 \\
    \textbf{Null Reference} & Attempt to access an attribute or method of a null object & 8     & 4 \\
    \textbf{Type Mismatch} & Using an incorrect data type in an operation or function call & 6     & 27 \\
    \textbf{Resource Limit Exceeded} & Exceeding the system's resource limits & 2     & 1 \\
    \textbf{Syntax error} & Incorrect syntax or structure in the code & 4     & 0 \\
    \textbf{Undefined Variable} & Accessing or using a variable that has not been defined & 8     & 6 \\
    \textbf{Attribute Not Found} & Attempt to access a non-existent attribute or method of an object & 3     & 7 \\
    \textbf{Duplicate Variable} & Defining a variable more than once in the same scope & 4     & 0 \\
    \bottomrule
    \end{tabular}%
    }
  \label{tab:rq2_bug_details}%
\end{table*}%

\subsubsection{Analysis on Compilation \& Runtime Errors}
Table~\ref{tab:rq2_bug_details} presents a comparison of common compilation and runtime errors error categories in Java and Python programs (i.e., 80 Python and 97 Java programs with the errors). 
From this table, we can observe that ChatGPT generates code containing a diverse range of errors across multiple categories, indicating the need for improvement in various aspects of code generation. 
Additionally, a significant portion of common compilation and runtime errors are relevant to the semantics of the generated program.
For example, these errors may contain illegal values (e.g., division by zero or invalid indices) and wrong access (e.g., concurrent modification, null references, and empty collection access). 
These observations can be explained by the probabilistic nature of the ChatGPT model, which predicts subsequent tokens based on preceding ones.  
This nature enables ChatGPT to understand the semantics of common programs that appear frequently in the training set.
However, the model captures the semantics implicitly from the training data, leading to misunderstandings of program semantics and subsequently resulting in semantically-related compilation and runtime errors. 
These findings indicate that incorporating semantic information into ChatGPT could potentially improve the quality of the generated code, indicating a promising direction for future research.
\find{
\textbf{Finding 5:} ChatGPT-generated code contains various types of execution errors, primarily due to misunderstandings of program semantics. 
}

We also notice that Illegal Index errors are quite prevalent in both languages, particularly in Java.
In fact, out of the 97 compilation and runtime errors encountered in Java, 45 of them (46.4\%) are attributed to using an invalid index. 
Meanwhile, Type Mismatch errors are more prevalent in Python than in Java, with 27 occurrences in Python compared to 6 in Java.
This observation could be due to Python's dynamic typing system, which allows for more flexibility in variable types, but can also lead to unexpected type-related issues at runtime.
Overall, these findings suggest that different languages may have distinct compilation and runtime error patterns and that improvements in code generation should take these language-specific characteristics into account. 
Additionally, the presence of various errors highlights the need for more effective debugging and error detection tools tailored to each language, ultimately leading to more robust and efficient code generation.

\find{
\textbf{Finding 6:} Different languages may have distinct compilation and runtime error patterns.
}

\begin{table*}[t]\centering
\caption{Top 10 Issues Affecting Code Style and Maintainability in Python Programs Generated by ChatGPT}\label{tab:rq2_python_smell}
\scalebox{0.59}{
\begin{tabular}{llccr}\toprule
\textbf{Errors} &\textbf{Descriptions} &\textbf{Pylint} &\textbf{Flake8} &\textbf{\#Programs}\\
\midrule
ConsiderUsingEnumerate &Used when code that iterates with range and len is encountered. &x & &213\\
NoElseReturn &Used in order to highlight an unnecessary block of code following an if containing a return statement. &x & &161\\
UnusedVariable &Used when a variable is defined but might not be used. &x &x &103\\
RedefinedBuiltin &Used when a variable or function override a built-in. &x & &63\\
ConsiderUsingDictItems &Used when iterating over the keys of a dictionary and accessing the value by index lookup. &x & &39\\
AvoidAmbigousNames &Used when code use variables named 'I', 'O', or 'l' & &x &38\\
TooManyBranches &Used when a function or method has too many branches, making it hard to follow. &x & &36\\
TooManyLocals &Used when a function or method has too many local variables. &x & &32\\
BlankLines &Nested functions should contain 1 blank line between their definitions. & &x &28\\
InconsistentReturnStatements &Either all return statements in a function should return an expression, or none of them should. &x & &27\\
\bottomrule
\end{tabular}}
\end{table*}
\begin{table*}[t]\centering
\caption{Top 10 Issues Affecting Code Style and Maintainability in Java Programs Generated by ChatGPT}
\label{tab:rq2_java_smell}
\scalebox{0.58}{
\begin{tabular}{llccr}\toprule
\textbf{Errors} &\textbf{Descriptions} &\textbf{CheckSyle} &\textbf{PMD} &\textbf{\#Programs}\\
\midrule
MultipleVariableDeclarations &Each variable declaration must be in its own statement &x & &334 \\
AvoidReassigningParameters &Emitted when incoming parameters are reassigned values & &x &176 \\
ForLoopCanBeForeach &Used to recommend to use foreach loop instead of loop. & &x &114 \\
RedundantModifier &Emitted when a modifier is redundant &x & &112 \\
RightCurly &Emitted when right curly in a code violate common conventions &x & &87 \\
VisibilityModifier & Used to recommend that a variable should not public &x & &86 \\
NPathComplexity &Used when a method have too many acyclic execution paths & &x &81 \\
LooseCoupling &Used when using implementation types instead of interface & &x &64 \\
HiddenField &Emitted when a local variable or a parameter does not shadow a field that is defined in the same class. &x & &55 \\
UseConcurrentHashMap &Recommend to use the ConcurrentHashMap implementation & &x &54 \\
\bottomrule
\end{tabular}}
\end{table*}

\subsubsection{Analysis on Code Style \& Maintainability}
Tables~\ref{tab:rq2_python_smell} and~\ref{tab:rq2_java_smell} present the top 10 issues affecting code style and maintainability in Python and Java programs generated by ChatGPT, respectively. From these tables, we can see various types of code styles and maintainability issues in the ChatGPT-generated code. 

In Python, the top three issues are ConsiderUsingEnumerate (213 out of 2033 programs, 10.5\%), NoElseReturn (161 out of 2033 programs, 7.9\%), and UnusedVariable (103 out of 2033 programs, 5.1\%). Interestingly, 5.1\% of ChatGPT-generated code has unused variables, which is considered a bad smell in code quality. Meanwhile, MultipleVariableDeclarations, AvoidingReassigningParameters, ForLoopCanBeForEach, and RedundantModifier are the most frequent issues happening in Java code generated by ChatGPT, accounting for more than 36\% ((334+176+114+112)/2033) of the generated code. The presence of these issues indicates that the code quality of ChatGPT-generated code for both Python and Java is not perfect and could be improved. 

We further compare code style and maintainability issues in Java and Python. Our results show that there are no overlapping top-10 issues in Python and Java. The possible reason is that Python and Java have very different code styles and common practices. These results highlight the need for language-specific techniques to address the issues. Finally, by analyzing the issues detected by different static analysis tools, we can see that there is only one common issue in Python that can be detected by both Pylint and Flake8. Similarly, there is no overlap between CheckStyle and PMD. Thus, using multiple static analysis tools can provide a more comprehensive analysis of code style and maintainability in ChatGPT-generated code.

\find{
\textbf{Finding 7:} ChatGPT-generated code contains various types of code style and maintainability issues. Their common issues are specific to the language and tool being used.  
}


\section{RQ3: Repair with Prompting}
\label{sec:RQ3}

\subsection{Repairing Prompt Types}
\smallsection{Experimental Design}
Sections~\ref{sec:RQ1} and~\ref{sec:RQ2} have demonstrated that ChatGPT is capable of generating functional code for various code generation tasks. 
However, the generated code sometimes suffers from different code quality issues, such as execution errors, wrong outputs, and maintainability problems.
Addressing these issues is vital to ensure the reliability and efficiency of the generated solutions.
Unlike traditional code generation tools, ChatGPT has the potential to learn from user interactions and refine its outputs based on the feedback it receives. 
This interactive process can lead to more accurate and high-quality code generation.
In this section, we investigate the self-repairing capabilities of ChatGPT in addressing the code quality issues identified in the generated code. 
We focus on providing effective feedback and exploring various strategies to enhance the performance of the model.
To investigate the impact of user feedback on the code quality of ChatGPT-generated solutions, we employ two types of feedback: (1) Simple Feedback and (2) Feedback with Static Analysis. 

\textbf{Simple Feedback:} This type of feedback involves providing ChatGPT with basic information about the issues in the generated code. For example, if a code quality issue is detected in the code, we provide feedback to ChatGPT as follows: "\textit{The generated code has quality issues. Please provide a better code implementation as expected by the task description.}"

\textbf{Feedback with Static Analysis and Runtime Errors:} In this method, we utilize the insights from static analysis tools and runtime errors (as discussed in Section~\ref{sec:RQ2}) to offer more precise and detailed feedback to ChatGPT. Thus, we augment the simple feedback with additional information derived from static analysis reports and runtime error messages. For example, if a static analysis tool pinpoints a specific error or poor coding practice, we supply ChatGPT with feedback that directly addresses the particular issue as follows: "\textit{The generated code contains the following quality issues: + $<$details from static analysis tools$>$ + Please provide a better code implementation as expected by the task description.}"

We use both types of feedback to prompt ChatGPT to refine and improve its generated code. Then, we compare the revised code with the original version to evaluate the effectiveness of the feedback in addressing the identified code quality issues.

\begin{figure}[t]
    \centering
    \includegraphics[width=0.9\textwidth]{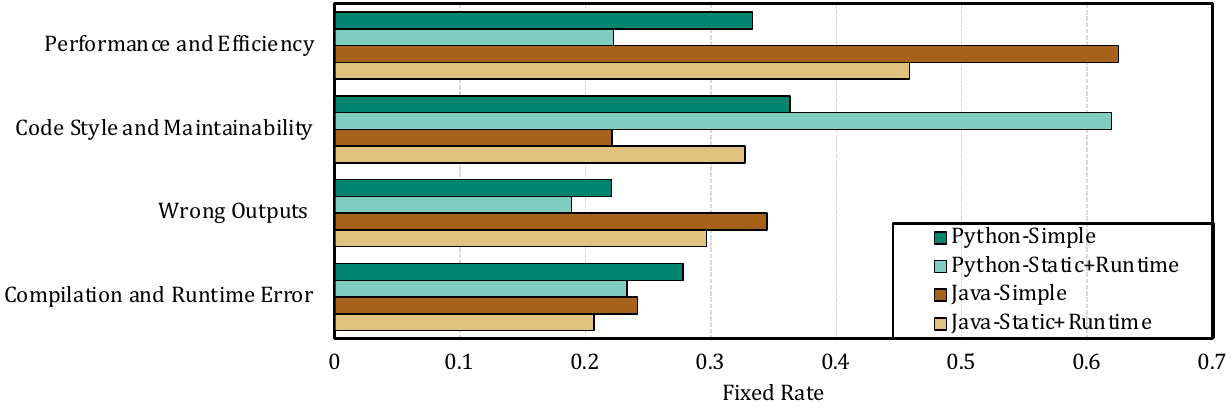}
    \caption{Comparison of Fix Rates for Different Feedback Types and Code Quality Issues}
    \label{fig:rq3_fix_rates}
\end{figure}

\smallsection{Result}
Figure~\ref{fig:rq3_fix_rates} presents the fixed rates for different feedback types and code quality issues for both Java and Python. 
The fixed rate is defined as the proportion of code quality issues that were successfully addressed and resolved by ChatGPT if the issue no longer happens, measured as a percentage (i.e., $\text{Fix Rate} = \frac{\text{Number of Issues Resolved}}{\text{Total Number of Issues}} $).
Overall, Figure~\ref{fig:rq3_fix_rates} presents that ChatGPT can successfully repair from about 20\% to 60\% code quality issues itself. Particularly, ChatGPT can resolve more than 60\% code style and maintainability issues in Python code with feedback from static analysis and runtime errors while more than 60\% performance and efficiency issues in Java code can be addressed with a simple feedback.

\find{
\textbf{Finding 8:} 
ChatGPT shows great promise in self-repairing code quality issues, achieving a fixed rate of 20\% to 60\%.
}

In our comparison of two prompt designs, we observed that feedback with static analysis and runtime errors is more effective in fixing code style and maintainability while simple feedback performs better in the remaining quality issues in both Java and Python. 
This is because feedback from static analysis tools provides detailed information about code quality issues, guiding ChatGPT to self-repairing these problems. 
For example, static analysis tools raise a warning that 
\begin{lstlisting}
    Solution.java:12: ForLoopCanBeForeach: This for loop can be replaced by a foreach loop
\end{lstlisting}
for the initial solution in lines 1 in Code~\ref{code:code-smell}. The warning provides detailed information about the code style and maintainability issue in line 12 including location and even solution. Therefore, ChatGPT can easily mitigate the issue. 
Meanwhile, feedback with runtime errors for remaining issues, such as execution errors or performance and efficiency, tends to be less specific and more ambiguous.
For example, in most of the performance and efficiency, we only obtain a ``TIMEOUT" message, which does not reveal any details or root cause of a given issue. Similarly, for solution inaccuracies, the runtime errors also usually only contain an \texttt{AssertionError}. For example, in Code~\ref{code:incorrect_example}, ChatGPT has only received the following information from runtime errors:
\begin{lstlisting}
    AssertionError : Input : cost = [10 , _15_ ,20] Expected output : 15
\end{lstlisting}
Although the \texttt{AssertionError} points out the incorrect input-output examples, it remains abstract and does not provide precise guidance. As a result of such limited feedback, it is not surprising that ChatGPT shows lower performance in self-repairing issues. Interestingly, we found that simple feedback is more effective than static analysis feedback or runtime errors in resolving these issues. 
This is possibly due to the introduction of noise by static analysis and runtime error feedback, which can confuse ChatGPT and lead to incorrect patches.
\find{
\textbf{Finding 9:} 
Prompts with detailed feedback can effectively assist ChatGPT in self-repairing code quality issues, whereas ambiguous feedback may have a negative impact on ChatGPT's performance.
}

\begin{figure}[t]
    \centering
    \includegraphics[width=0.9\textwidth]{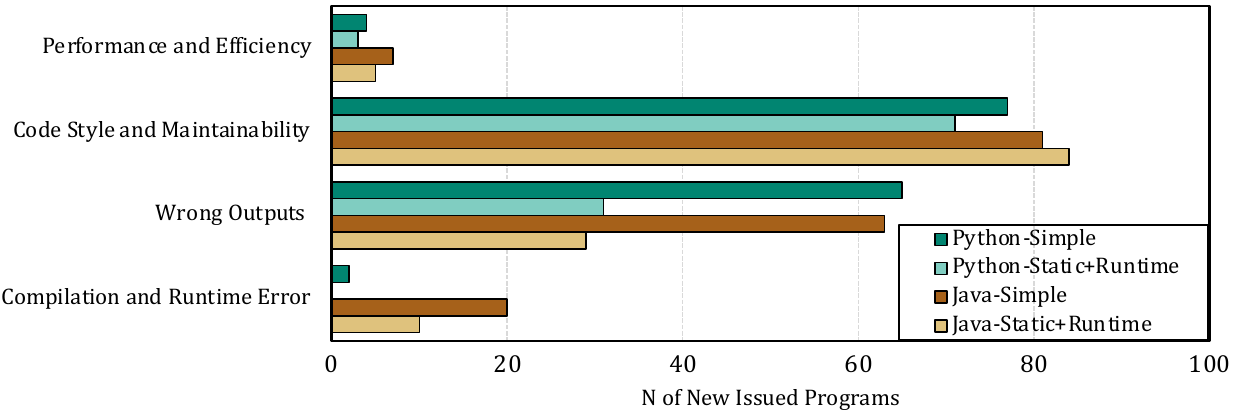}
    \caption{Number of Newly Introduced Quality Issues after Providing Feedback}
    \label{fig:rq3_new_isssues_during_fixing}
\end{figure}

While the aforementioned feedback are quite effective in guiding ChatGPT to self-mitigate its code quality issues, we still observe that new code quality issues are introduced in the fixed version. Figure~\ref{fig:rq3_new_isssues_during_fixing} presents the number of newly introduced quality issues after providing feedback for both Java and Python. 
The results demonstrate that, in some cases, addressing certain code quality issues may inadvertently introduce new issues in the generated code. 
It can be seen that providing simple feedback results in a higher number of new issues compared to static feedback and runtime errors. 
This is because simple feedback is quite general so ChatGPT is sometimes misguided, resulting in introducing new problems though plausibly solving the original issues. We observed that ChatGPT sometimes even generates a new code, which is totally different from the original version. 
\begin{lstlisting}[language=java, caption=A example of newly-introduced issues in ChatGPT-generated fix using simple feedback for LeetCode Problem 1323-maximum-69-number, label=code:incorrect_simple_fix_example]
class Solution {
    //Initial solution with issues
    public int maximum69Number (int num) {
        int[] arr = String.valueOf(num).chars().map(c -> c-'0').toArray();
        for(int i=0; i<arr.length; i++){
            if(arr[i] == 6){
                arr[i] = 9;
                break;
            }
        }
        int res = 0;
        for(int i=0; i<arr.length; i++){
            res = res*10 + arr[i];
        }
        return res;
    }
    //Fixed by simple feedback
    public int maximum69Number (int num) {
        int digits = (int) Math.log10(num) + 1;
        int firstSixDigit = digits - 1;
        int temp = num;
        while (temp > 0) {
            if (temp % 10 == 6) {
                break;
            }
            temp = temp / 10;
            firstSixDigit--;
        }
        return num + (int) Math.pow(10, firstSixDigit) * 3;
    }
    //Fixed by static analysis feedback
    public int maximum69Number (int num) {
        int[] arr = Integer.toString(num).chars().map(c -> c - '0').toArray();
        for (int i = 0; i < arr.length; i++) {
            if (arr[i] == 6) {
                arr[i] = 9;
                break;
            }
        }
        int res = 0;
        for (int digit : arr) {
            res = res * 10 + digit;
        }
        return res;
    }
}
\end{lstlisting}
For example, lines 18-30 in Code~\ref{code:incorrect_simple_fix_example} show a fix generated by ChatGPT for the initial solution in lines 3-16. Unfortunately, instead of fixing the issue, ChatGPT generated a new solution (lines 18-30), which implement an incorrect solution, resulting in failing test cases. Static feedback and runtime errors, on the other hand, provide detailed information, i.e., leading to a correct fix in line 41 which change the for-loop to foreach-loop. The results show that providing more accurate feedback about code quality issues could lead to improvement in the quality of fixed programs by ChatGPT. These findings emphasize the importance of advanced feedback mechanisms and strategies that improve ChatGPT's self-repairing capabilities by reducing the introduction of new issues while resolving existing code quality problems.
\find{
\textbf{Finding 10:} Despite being effective in self-repairing code quality issues, ChatGPT still introduces new code quality issues in the generated fixes. More precise feedback could help mitigate the issues. 
}

\begin{figure}[t]
  \centering
  \begin{minipage}{0.49\linewidth}
    \centering
    \includegraphics[width=\linewidth, height=4cm, keepaspectratio]{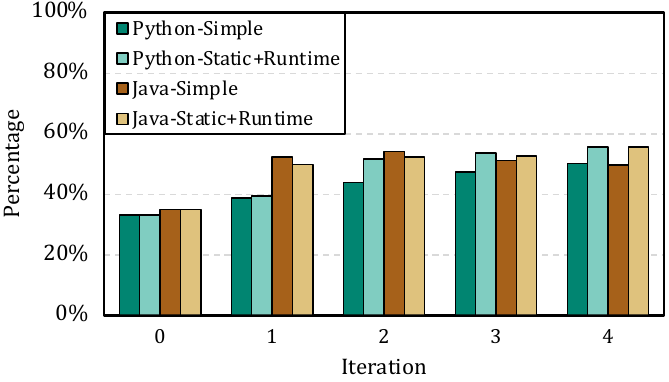} 
    \caption{Pass Rates Across Iterative Feedback Rounds on 402 Tasks after 2022}
    \label{fig:rq4_iteration_pass_rate}
  \end{minipage}%
  \hfill
  \begin{minipage}{0.49\linewidth} 
    \centering
    \includegraphics[width=\linewidth, height=4cm, keepaspectratio]{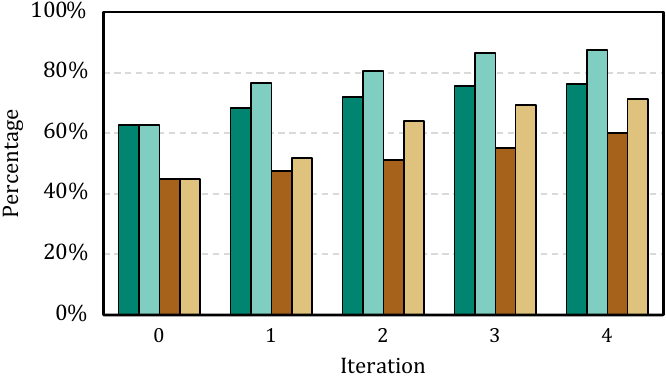}
    \caption{Percentage of Code without Code Style and Maintainability Issues Across Iterations}
    \label{fig:rq4_code_smell}
  \end{minipage}
\end{figure}

\subsection{Iterative Repairing}
In previous evaluations, we assessed the self-repairing capabilities of ChatGPT by providing individual feedback for each identified code quality issue. 
However, in a realistic setting, users may interact with ChatGPT iteratively, providing successive feedback until the generated code meets their satisfaction. 
This subsection, therefore, explores the performance of ChatGPT in iterative repairing scenarios involving multiple rounds of feedback.

\smallsection{Experimental Design}
We use the same two types of feedback as in the previous subsection: simple feedback and feedback with static analysis and runtime errors. 
To make the experiment more rigorous, we only use the 402 new test examples from LeetCode that were published in the year 2022 (after GPT-3.5’s pre-training knowledge cutoff).
In each iteration, we provide feedback based on the code quality issues identified in the last iteration’s generated code.
The feedback is provided interactively, simulating a real-world conversation with ChatGPT.
If the generated code has no quality issues, we end the iteration.
In our experiments, we conduct four iterative rounds to evaluate the quality of the code generated by ChatGPT.

\begin{figure}[t]
    \centering
    \includegraphics[width=0.55\textwidth]{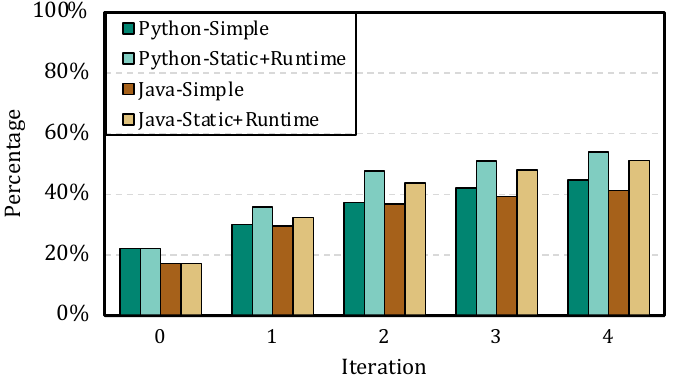}
    \caption{Iterative Feedback Impact on Producing Code Without Quality Issues}
    \label{fig:rq4_overall_quality}
\end{figure}

\smallsection{Result}
Figures~\ref{fig:rq4_iteration_pass_rate}, ~\ref{fig:rq4_code_smell}, and ~\ref{fig:rq4_overall_quality} present our iterative repairing results on 402 new LeetCode tasks released after 2022.
The x-axis represents the iteration count, with iteration 0 corresponding to the original code generated by ChatGPT in response to the initial code generation prompt. 
Subsequent iterations, labeled 1 to 4, represent the feedback rounds provided to ChatGPT.
As shown in Figure~\ref{fig:rq4_iteration_pass_rate}, for 402 new test examples, overall, the pass rate shows a clear increase in iteration round 1 and 2, and then becomes stable.
This demonstrates ChatGPT’s ability to integrate feedback to improve functional correctness.
Interestingly, for Java code generation using simple feedback, there is a small decline in the pass rate compared to other scenarios. This finding is caused by the fact that simple feedback only indicates that the generated code has issues, without providing specific details.
Figure~\ref{fig:rq4_code_smell} presents the percentage of generated code without code style and maintainability issues. 
There is a stable increase across all four scenarios. 
It is also consistent with Figure~\ref{fig:rq3_fix_rates} that feedback with static analysis and runtime errors results in a higher rate of issue resolution compared to simple feedback.
Figure~\ref{fig:rq4_overall_quality} shows the overall percentage of code without any of the different quality issues listed in Section~\ref{sec:RQ2} (i.e., no errors, no code smells).
It is clear that iterative repairing is effective in improving code quality.
Moreover, although our previous experiments indicated that simple feedback may be better for some issues, iterative feedback with static analysis and runtime errors performs much better than simple feedback as the iteration rounds progress.

\find{
\textbf{Finding 11:} Iterative repairing proves to be effective, particularly when guided by detailed feedback that incorporates static analysis and runtime errors.
}




\section{Discussion}
\label{sec:discussion}

\subsection{Enhancing ChatGPT’s Code Generation and Self-Repair Capabilities}
In this subsection, we delve into strategies to potentially enhance ChatGPT's code generation and self-repair capabilities in the real-world scenario.
\subsubsection{Prompt Optimization}
Recent research highlights the crucial role of prompt engineering in enhancing the performance of large language models (LLMs) like ChatGPT for software engineering tasks~\cite{hou2023large}.
This process, which involves the careful design of specialized prompts, is a fundamental technique for improving interactions with LLMs such as ChatGPT~\cite{dong2023self}.
For example, Gao~\ea~\cite{gao2023constructing} demonstrated that incorporating additional examples into the prompt could potentially enhance performance in bug fixing tasks.
Meanwhile, Ahmed~\ea~\cite{ahmed2022few} showed that in code summarization tasks, augmenting a prompt with explicit semantic facts can significantly improve performance.
Our findings, presented in Section~\ref{sec:RQ3}, corroborate these previous studies, indicating that the effectiveness of ChatGPT in self-repairing code issues is significantly influenced by the quality and specificity of prompts.
We found that for code style and maintainability issues (where static analysis tools like Pylint provide precise guidance including location and even solution), prompts that self-repair using static and runtime information achieve better performance than simple prompts that merely identify a quality issue.
However, for errors related to wrong outputs or efficiency, simpler prompts worked better in single-round feedback, likely because the prompts collected from the compiler feedback might be ambiguous or unhelpful.
These findings suggest that highly specific and well-crafted prompts can significantly enhance the performance of LLMs like ChatGPT in software engineering contexts.
Therefore, this highlights the importance of continued research in prompt engineering to fully harness the potential of LLMs in software engineering.

\subsubsection{Iterative Interactions}
In Section~\ref{sec:RQ3}, we demonstrated that iterative repairing is effective, with code quality improving as the iteration rounds progress.
In real-world usage, interactions with ChatGPT can be iterative, whereby the user provides feedback or additional information to ChatGPT after each prompt.
Xia et al.~\cite{xia2023conversational} observed that ChatGPT’s performance in generating correct patches improves notably as the number of iterations increases, with a significant improvement observed around three iterations.
Our research further proves that feedback from detailed static analysis tools and compilers can effectively enhance ChatGPT’s code repair capability over iterative interactions.
This evidence spotlights how repeated interactions enable ChatGPT to refine its understanding, adjust based on user feedback, and converge towards more accurate solutions.
However, our results show performance stabilizing in later rounds, indicating potential upper bounds to iterative gains.
Therefore, future work should establish interaction design patterns and benchmarks to systematically advance the efficiency and efficacy of conversational code generation.

\subsection{Lessons Learned}
In this section, we highlight key lessons learned through our experiments and analysis that can drive future research in the field.

\textbf{Code quality issues are prevalent in AI-generated code:} Our study revealed that ChatGPT-generated code is prone to various code quality issues, including compilation and runtime errors, wrong outputs, and maintainability problems.
This finding emphasizes the importance of addressing these issues to ensure the long-term success of AI-driven code generation and to maintain high-quality software systems.

\textbf{Task difficulty, time that tasks are introduced, and program size impact automated code generation performance:} We found that the performance of ChatGPT on code generation tasks is significantly influenced by factors such as task difficulty, task-established time, and program size. This suggests that improvements in AI models should consider these factors to better adapt to different types of code generation tasks.

\textbf{Tailored feedback and prompt engineering are crucial for effective self-repairing and code generation quality:} Our results suggest that the effectiveness of ChatGPT's self-mitigating capabilities depends on the type of feedback provided, the programming language, and the specific code quality issue. For instance, static analysis feedback works better for code style and maintainability issues, while simple feedback is more effective for addressing execution errors and wrong outputs. This finding highlights the importance of providing tailored feedback to maximize the efficacy of ChatGPT's self-mitigating capabilities. Moreover, the quality of ChatGPT-generated code can be heavily affected by the choice of prompts. Future work could explore optimizing prompts to improve the accuracy and reliability of ChatGPT-generated code, further enhancing the overall effectiveness of AI-driven code generation models.


\subsection{Threats to Validity}
Below, we discuss threats that may impact the results of our study.

\subsubsection{External validity} 
Threats to external validity concern the generalizability of our findings. 
Our study is based on a dataset of 2,033 programming tasks from LeetCode, which may not represent all possible code generation tasks encountered in real-world software development. 
Additionally, we focus on Java and Python, two popular programming languages; however, our findings may not be directly applicable to other programming languages. 
To mitigate these threats, future work could expand the dataset by incorporating tasks from various sources and diverse programming languages, and by considering different types of software projects, such as web applications, mobile apps, and embedded systems.

\subsubsection{Internal validity}
Threats to internal validity refer to possible errors in our experiments. One such threat relates to bugs happening in our code. To mitigate this risk, we have carefully checked our code and made our code publicly available~\cite{replication}. Another possible threat may be introduced from our manual analysis and categorization. To eliminate the potential bias, we conducted a sorting discussion among three annotators. We also release our analysis and categorization results for public verification.   
In addition, to minimize the non-deterministic nature of ChatGPT, we set the temperature parameter to 0 in our experiments. This approach ensures that ChatGPT produces consistent outputs for the same input, thereby reducing variability and enhancing the internal validity of our results.

\subsubsection{Construct validity} 
Threats to construct validity relate to the suitability of our evaluation. In our study, we use the pass@1 metric, in which a program is considered as functionally correct if it passes all the test cases. A possible threat arises from the incompleteness of the test suite, which could potentially result in missed program bugs. In our experiments, we use the original test suite from LeetCode, which is carefully designed and widely recognized. Thus, we believe this risk is minimal. Another potential threat to construct validity comes from the variability in ChatGPT-generated code due to different prompts. To address this concern, we followed the similar methodology used by Fan~\ea~\cite{fan2022automated} and Tian~\ea~\cite{tian2023chatgpt}, which ensures that our results are reliable by using a consistent set of prompts across different tasks. However, it is important to note that prompt engineering can significantly influence the quality of the generated code. Future work could focus on optimizing the prompts to improve the accuracy and reliability of ChatGPT-generated code, thus enhancing the overall effectiveness of AI-driven code generation models.


\section{Related work}
\label{sec:relatedwork}
In this section, we present related work and discuss the novelty of our work with respect to large language models for code generation and  code quality issues. 

\subsection{Large Language Model for Code Generation}
Large Language Models have been emerging as the state-of-the-art in code-related tasks, advancing the progress on program understanding~\cite{feng2020codebert, wang2021codet5, ahmed2022few, le2023invalidator}, analysis~\cite{kazerounian2021simtyper, le2022autopruner, thapa2022transformer, nguyen2023multi}, and generation~\cite{chen2021evaluating, li2022competition, jain2022jigsaw, xia2023automated}. Among these tasks, LLM-based code generation approaches are closely related to our study. The era of LLMs in code generation began with the introduction of CodeX~\cite{chen2021evaluating}, which serves as the backend for a well-known commercial tool, i.e., GitHub Copilot~\cite{githubcopilot}.
After the success of CodeX, various models such as InCoder~\cite{fried2022incoder}, Google Alphacode~\cite{li2022competition}, and Amazon CodeWhisperer~\cite{amazoncodewhisperer} have been emerging, resulting in remarkable improvements in the effectiveness of code generation. These advancements provide a new way of tackling the code generation problem. Recently, OpenAI introduced ChatGPT~\cite{chatgpt} a general AI-powered chatbot with remarkable capabilities in language understanding and human-like answering. ChatGPT has shown remarkable accuracy in generating code and solving programming problems while receiving positive feedback from users and gaining popularity. 
However, the quality of the ChatGPT-generated code is the critical concern for top software companies when deciding whether to adopt it in practice or not~\cite{chatgptconcern1, chatgptconcern2, chatgptconcern3}.

To the best of our knowledge, this paper is the first to conduct a time-sensitive evaluation of ChatGPT on a code generation task. Moreover, it is the first to systematically analyze and characterize the code quality issues in ChatGPT-generated code and explore potential solutions to repair them. By doing so, we hope to increase awareness about the quality issues in code generated by ChatGPT, and provide suggestions for mitigating the issues. 

\subsection{Code Quality Issues}

Code quality issues are the most important concern, as one quality issue (aka. software defect) could lead to monetary and reputation costs.
There is a large number of studies investigating code quality issues from human-written code.
For example, Kochhar~\ea~\cite{kochhar2016large} conducted a large-scale empirical investigation on the code quality of open-source projects implemented in 17 programming languages. 
Saboury~\ea~\cite{saboury2017empirical} empirically investigate code smells in 537 releases of five popular Javascript applications. 
Keuning~\ea~\cite{keuning2017code} investigate code quality issues in student programs. 

However, none of these studies focuses on the code quality issues for the AI-generated code, highlighting the difference between our paper and the literature on code quality issues.

\subsection{Code Quality Issues of AI-generated Code}

As AI-generated code becomes more prominent, researchers investigate the quality of code generated by CodeX and GitHub Copilot. 
For example, 
Nguyen~\ea~\cite{nguyen2022empirical} evaluated the performance of Copilot on 33 programming tasks. 
Fan~\ea~\cite{fan2022automated} further analyze common bugs in code generated by Codex, the backend of Copilot, on 113 programming tasks and benchmark automated program repair tools on fixing the mistakes.
However, these studies focus on CodeX and GitHub Copilot on a few programming tasks, leading to a lack of diversity. 
Different from these studies, we conduct a large-scale analysis on 2033 programming tasks, which enables us not only to comprehensively evaluate the effectiveness of AI-generated code but also to identify factors affecting their performance. 
Moreover, Nguyen~\ea~\cite{nguyen2022empirical} focus on analyzing the correctness of AI-generated code while Fan~\ea~\cite{fan2022automated} target to benchmark automated program repair tools on fixing mistakes. 
Our work, on the other hand, delves deeper into analyzing code quality issues, including code style \& maintainability issues, and highlights common patterns across different types of code quality problems.
Besides that, our study focuses on ChatGPT, a recently-introduced AI chatbot developed by OpenAI. Unlike Codex and GitHub Copilot, which target experienced professional developers and programmers, ChatGPT caters to a much wider audience, with approximately 1.6 billion visits in April 2023~\cite{chatgpttraffic}, including novice programmers and non-coders. Therefore, a comprehensive study on the reliability of source code generated by ChatGPT would not only provide valuable insights into the model but also raise awareness about the responsible usage of ChatGPT in code generation.

\section{Conclusion}
\label{sec:conclusion}

In this study, we conducted a systematic analysis of ChatGPT-generated code to assess its reliability and identify potential code quality issues.
Our findings demonstrate that while ChatGPT can generate functional code for various programming tasks, the generated code often suffers from quality issues, such as compilation and errors, wrong outputs, maintainability problems, and performance inefficiencies.
We also explored ChatGPT's self-repairing capabilities and investigated the impact of different feedback strategies in addressing these code quality issues.

Our research provides valuable insights into the current limitations of ChatGPT and highlights the importance of considering context-aware feedback and code quality issues when utilizing AI-driven code generation tools. 
Moreover, our work offers insights for future research and development efforts aimed at enhancing the code generation capabilities of AI models like ChatGPT. 
We believe that by addressing these challenges, we can pave the way for more reliable, efficient, and maintainable AI-generated code, ultimately benefiting both experienced developers and novice programmers alike. In the future, we plan to develop more advanced prompts in both code generation and fixing to further improve the reliability of ChatGPT-generated code.


\bibliographystyle{ACM-Reference-Format}
\bibliography{sample-base}

\end{document}